\documentclass[aps,prl,twocolumn,superscriptaddress,showpacs,preprintnumbers,amsmath,amssymb]{revtex4}
\usepackage{color}

\usepackage{graphicx} 
\usepackage{dcolumn}  
\RequirePackage{xspace}
\usepackage{relsize}
\def\babar{\mbox{\slshape B\kern-0.1em{\smaller A}\kern-0.1em B\kern-0.1em{\smaller A\kern-0.2em R}}}

\def\qqbar {\ensuremath{q\overline q}\xspace}
\def\piz   {\ensuremath{\pi^0}\xspace}
\def\pip   {\ensuremath{\pi^+}\xspace}
\def\pim   {\ensuremath{\pi^-}\xspace}
\def\Kbar  {\kern 0.2em\overline{\kern -0.2em K}{}\xspace}

\def\Kz    {\ensuremath{K^0}\xspace}
\def\Kzb   {\ensuremath{\Kbar^0}\xspace}
\def\Kp    {\ensuremath{K^+}\xspace}
\def\Km    {\ensuremath{K^-}\xspace}
\def\Kmp   {\ensuremath{K^{\mp}}\xspace}
\def\KS    {\ensuremath{K^0_{\scriptscriptstyle S}}\xspace} 

\def\Kstar {\ensuremath{K^*}\xspace}

\def\Dbar  {\kern 0.2em\overline{\kern -0.2em D}{}\xspace}

\def\Dz    {\ensuremath{D^0}\xspace}
\def\Dzb   {\ensuremath{\Dbar^0}\xspace}

\def\Dstarp{\ensuremath{D^{*+}}\xspace}

\def\Bbar  {\kern 0.18em\overline{\kern -0.18em B}{}\xspace}

\def\BB    {\ensuremath{B\Bbar}\xspace} 
\def\Bz    {\ensuremath{B^0}\xspace}
\def\Bzb   {\ensuremath{\Bbar^0}\xspace}
\def\Bu    {\ensuremath{B^+}\xspace}
\def\Bub   {\ensuremath{B^-}\xspace}
\def\Bp    {\ensuremath{\Bu}\xspace}
\def\Bm    {\ensuremath{\Bub}\xspace}
\mathchardef\Upsilon="7107
\def\Y#1S{\ensuremath{\Upsilon{(#1S)}}\xspace}
\def\mbc   {\mbox{$M_{\rm bc}$}\xspace}
\def\DeltaE{\mbox{$\Delta E$}\xspace}
\def\cm    {\ensuremath{{\rm \,cm}}\xspace}
\def\invfb {\ensuremath{\mbox{\,fb}^{-1}}\xspace}
\def\order {{\ensuremath{\cal O}}\xspace}
\def\to    {\ensuremath{\rightarrow}\xspace}
\def\CP    {\ensuremath{C\!P}\xspace}
\def\etal  {{\it et~al.}}
\def\nb    {\ensuremath{C_{N\!B}}\xspace}
\def\nbprim{\ensuremath{C'_{N\!B}}\xspace}
\def\nbmin {\ensuremath{C_{N\!B,{\rm min}}}\xspace}
\def\nbmax {\ensuremath{C_{N\!B,{\rm max}}}\xspace}
\newcommand{\stat}{\ensuremath{\mathrm{(stat)}}\xspace}
\newcommand{\syst}{\ensuremath{\mathrm{(syst)}}\xspace}
\newcommand{\gev}{\ensuremath{\mathrm{\,Ge\kern -0.1em V}}\xspace}
\newcommand{\mev}{\ensuremath{\mathrm{\,Me\kern -0.1em V}}\xspace}
\newcommand{\gevc}{\ensuremath{{\mathrm{\,Ge\kern -0.1em V\!/}c}}\xspace}
\newcommand{\mevc}{\ensuremath{{\mathrm{\,Me\kern -0.1em V\!/}c}}\xspace}
\newcommand{\gevcc}{\ensuremath{{\mathrm{\,Ge\kern -0.1em V\!/}c^2}}\xspace}
\newcommand{\mevcc}{\ensuremath{{\mathrm{\,Me\kern -0.1em V\!/}c^2}}\xspace}

\begin{document}

\vspace*{-3\baselineskip}

\preprint{\vbox{ \hbox{   }
			\hbox{Belle Preprint {\it 2013-5}}
			\hbox{KEK Preprint {\it 2013-1}}
}}

\title{ \quad\\[1.0cm] Evidence for the decay {\boldmath $\Bz\to K^+K^-\pi^0$}}

\noaffiliation
\affiliation{University of the Basque Country UPV/EHU, 48080 Bilbao}
\affiliation{Budker Institute of Nuclear Physics SB RAS and Novosibirsk State University, Novosibirsk 630090}
\affiliation{Faculty of Mathematics and Physics, Charles University, 121 16 Prague}
\affiliation{University of Cincinnati, Cincinnati, Ohio 45221}
\affiliation{Deutsches Elektronen--Synchrotron, 22607 Hamburg}
\affiliation{Justus-Liebig-Universit\"at Gie\ss{}en, 35392 Gie\ss{}en}
\affiliation{Gifu University, Gifu 501-1193}
\affiliation{II. Physikalisches Institut, Georg-August-Universit\"at G\"ottingen, 37073 G\"ottingen}
\affiliation{Gyeongsang National University, Chinju 660-701}
\affiliation{Hanyang University, Seoul 133-791}
\affiliation{University of Hawaii, Honolulu, Hawaii 96822}
\affiliation{High Energy Accelerator Research Organization (KEK), Tsukuba 305-0801}
\affiliation{Ikerbasque, 48011 Bilbao}
\affiliation{Indian Institute of Technology Guwahati, Assam 781039}
\affiliation{Indian Institute of Technology Madras, Chennai 600036}
\affiliation{Institute of High Energy Physics, Chinese Academy of Sciences, Beijing 100049}
\affiliation{Institute of High Energy Physics, Vienna 1050}
\affiliation{Institute for High Energy Physics, Protvino 142281}
\affiliation{Institute of Mathematical Sciences, Chennai 600113}
\affiliation{INFN - Sezione di Torino, 10125 Torino}
\affiliation{Institute for Theoretical and Experimental Physics, Moscow 117218}
\affiliation{J. Stefan Institute, 1000 Ljubljana}
\affiliation{Kanagawa University, Yokohama 221-8686}
\affiliation{Institut f\"ur Experimentelle Kernphysik, Karlsruher Institut f\"ur Technologie, 76131 Karlsruhe}
\affiliation{Korea Institute of Science and Technology Information, Daejeon 305-806}
\affiliation{Korea University, Seoul 136-713}
\affiliation{Kyungpook National University, Daegu 702-701}
\affiliation{\'Ecole Polytechnique F\'ed\'erale de Lausanne (EPFL), Lausanne 1015}
\affiliation{Faculty of Mathematics and Physics, University of Ljubljana, 1000 Ljubljana}
\affiliation{Luther College, Decorah, Iowa 52101}
\affiliation{University of Maribor, 2000 Maribor}
\affiliation{Max-Planck-Institut f\"ur Physik, 80805 M\"unchen}
\affiliation{School of Physics, University of Melbourne, Victoria 3010}
\affiliation{Moscow Physical Engineering Institute, Moscow 115409}
\affiliation{Moscow Institute of Physics and Technology, Moscow Region 141700}
\affiliation{Graduate School of Science, Nagoya University, Nagoya 464-8602}
\affiliation{Kobayashi-Maskawa Institute, Nagoya University, Nagoya 464-8602}
\affiliation{Nara Women's University, Nara 630-8506}
\affiliation{National Central University, Chung-li 32054}
\affiliation{National United University, Miao Li 36003}
\affiliation{Department of Physics, National Taiwan University, Taipei 10617}
\affiliation{H. Niewodniczanski Institute of Nuclear Physics, Krakow 31-342}
\affiliation{Nippon Dental University, Niigata 951-8580}
\affiliation{Niigata University, Niigata 950-2181}
\affiliation{Osaka City University, Osaka 558-8585}
\affiliation{Pacific Northwest National Laboratory, Richland, Washington 99352}
\affiliation{Panjab University, Chandigarh 160014}
\affiliation{University of Pittsburgh, Pittsburgh, Pennsylvania 15260}
\affiliation{Punjab Agricultural University, Ludhiana 141004}
\affiliation{Research Center for Electron Photon Science, Tohoku University, Sendai 980-8578}
\affiliation{University of Science and Technology of China, Hefei 230026}
\affiliation{Seoul National University, Seoul 151-742}
\affiliation{Sungkyunkwan University, Suwon 440-746}
\affiliation{School of Physics, University of Sydney, NSW 2006}
\affiliation{Tata Institute of Fundamental Research, Mumbai 400005}
\affiliation{Excellence Cluster Universe, Technische Universit\"at M\"unchen, 85748 Garching}
\affiliation{Toho University, Funabashi 274-8510}
\affiliation{Tohoku Gakuin University, Tagajo 985-8537}
\affiliation{Tohoku University, Sendai 980-8578}
\affiliation{Department of Physics, University of Tokyo, Tokyo 113-0033}
\affiliation{Tokyo Institute of Technology, Tokyo 152-8550}
\affiliation{Tokyo Metropolitan University, Tokyo 192-0397}
\affiliation{Tokyo University of Agriculture and Technology, Tokyo 184-8588}
\affiliation{University of Torino, 10124 Torino}
\affiliation{CNP, Virginia Polytechnic Institute and State University, Blacksburg, Virginia 24061}
\affiliation{Wayne State University, Detroit, Michigan 48202}
\affiliation{Yamagata University, Yamagata 990-8560}
\affiliation{Yonsei University, Seoul 120-749}
  \author{V.~Gaur}\affiliation{Tata Institute of Fundamental Research, Mumbai 400005} 
  \author{G.~B.~Mohanty}\affiliation{Tata Institute of Fundamental Research, Mumbai 400005} 
  \author{T.~Aziz}\affiliation{Tata Institute of Fundamental Research, Mumbai 400005} 
  \author{I.~Adachi}\affiliation{High Energy Accelerator Research Organization (KEK), Tsukuba 305-0801} 
  \author{H.~Aihara}\affiliation{Department of Physics, University of Tokyo, Tokyo 113-0033} 
  \author{D.~M.~Asner}\affiliation{Pacific Northwest National Laboratory, Richland, Washington 99352} 
  \author{V.~Aulchenko}\affiliation{Budker Institute of Nuclear Physics SB RAS and Novosibirsk State University, Novosibirsk 630090} 
  \author{T.~Aushev}\affiliation{Institute for Theoretical and Experimental Physics, Moscow 117218} 
  \author{A.~M.~Bakich}\affiliation{School of Physics, University of Sydney, NSW 2006} 
  \author{A.~Bala}\affiliation{Panjab University, Chandigarh 160014} 
% ...
\collaboration{Belle Collaboration}

\begin{abstract}
We report a search for charmless hadronic decays of neutral $B$ mesons to the
final state $K^+K^-\pi^0$. The results are based on a $711\invfb$ data sample that
contains $772\times 10^6$ $\BB$ pairs, and was collected at the $\Y4S$ resonance
with the Belle detector at the KEKB asymmetric-energy $e^+e^-$ collider. We find
the first evidence for this decay with a significance of $3.5$ standard deviations
and measure its branching fraction as ${\cal B}(\Bz\to K^+K^-\pi^0)=[2.17\pm0.60
\stat\pm0.24\syst]\times10^{-6}$.
\end{abstract}


\maketitle

\tighten

{\renewcommand{\thefootnote}{\fnsymbol{footnote}}}
\setcounter{footnote}{0}

\begin{figure}[!htb]
\begin{center}$
\begin{array}{cc}
\includegraphics[width=.24\textwidth]{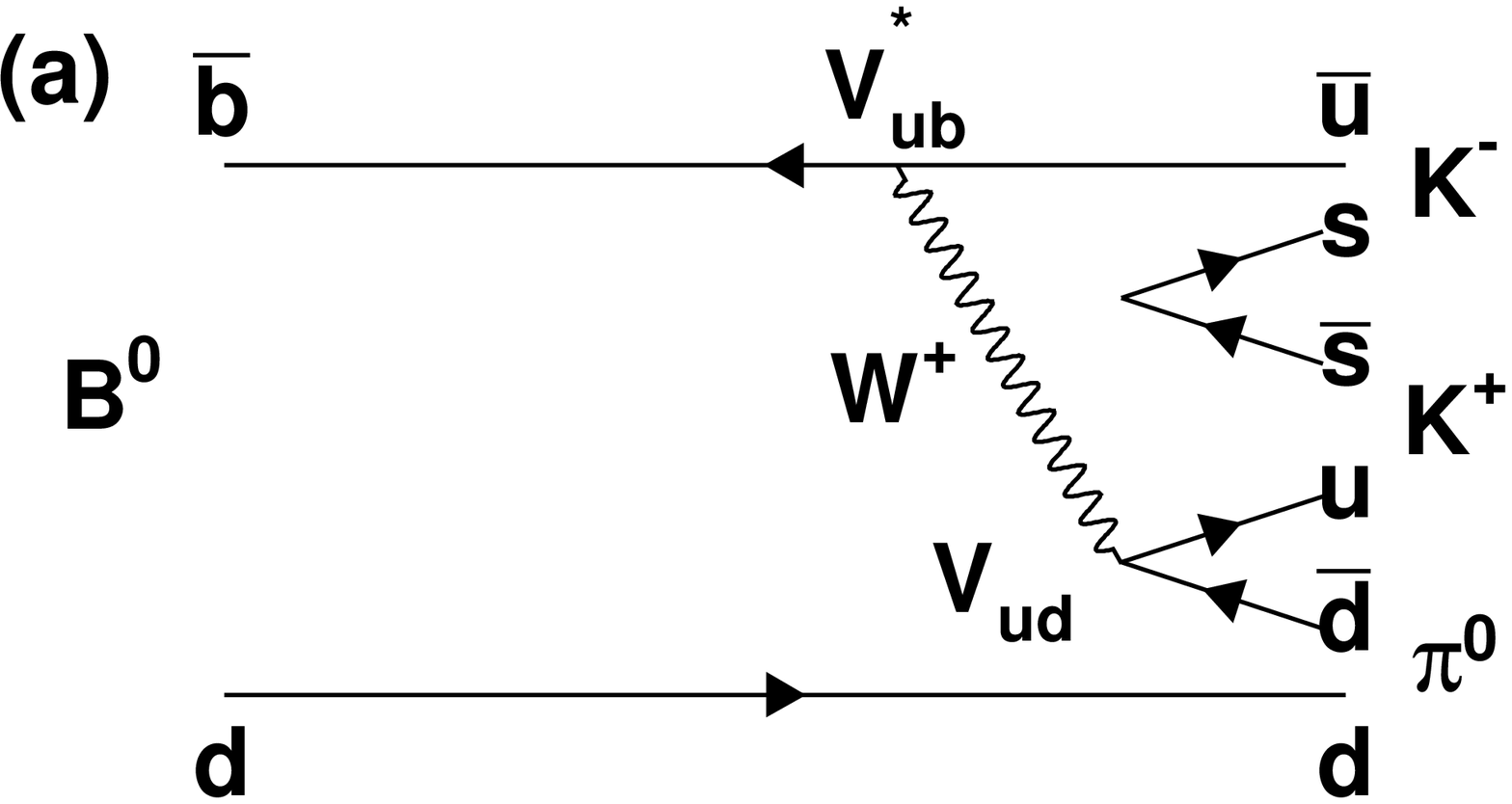} &
\includegraphics[width=.24\textwidth]{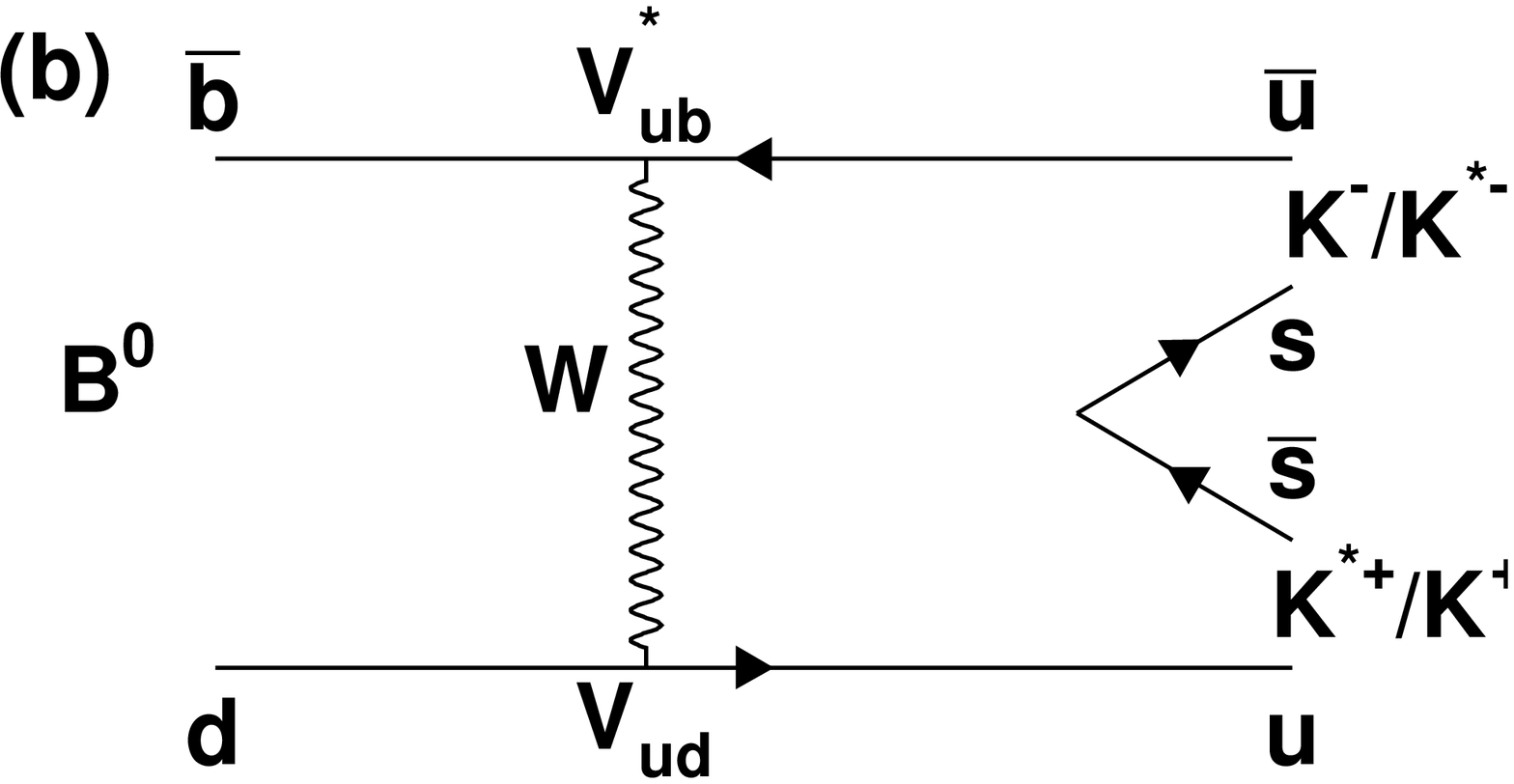} \\
\end{array}$
\end{center}
\caption{Typical Feynman diagrams that contribute to the
 decay $\Bz\to\Kp\Km\piz$: (a) $b\to u$ tree and (b) internal
 $W$ exchange.}
\label{fig:Feyn_diag}
\end{figure}

The $B$-meson decay $\Bz\to\Kp\Km\piz$ is suppressed in the
standard model (SM) and thus offers a useful probe for new
physics beyond the SM. Figure~\ref{fig:Feyn_diag} shows
typical Feynman diagrams that contribute to this decay. The
dominant one is the color- and Cabibbo-suppressed $b\to u$
tree transition followed by the internal $W$ exchange diagram
leading to $\Bz\to K^{*\pm}K^{\mp}$ with $K^{*\pm}\to K^{\pm}
\piz$. The latter diagram dominates in the decay $\Bz\to\Kp\Km$,
for which only upper limits have been placed on the branching
fraction~\cite{Aubert:2006fha,Aaltonen:2011jv,Aaij:2012as,Duh:2012ie}.
This is in contrast to the related decays (having two kaons
in the final state) that are already observed such as $\Bz\to
\Kz\Kzb$, $\Bp\to\Kz\Kp$~\cite{Duh:2012ie,Aubert:2006gm}, and
$\Bp\to\Kp\Km\pip$~\cite{Aubert:2007xb,lhcb}, where the $b\to d$
gluonic penguin amplitude can contribute as well~\cite{charge}.

The three-body decay $\Bz\to\Kp\Km\piz$ has not yet been observed,
with only one measured upper limit of ${\cal B}(\Bz\to\Kp\Km\piz)
<19\times 10^{-6}$ at $90\%$ confidence level from the CLEO
Collaboration~\cite{Eckhart:2002qr}. Intermediate resonant modes
that decay preferentially to this final state have also not been
seen. A search for a related channel by Belle has set an upper
limit of ${\cal B}(\Bz\to\phi\piz)<1.5\times10^{-7}$~\cite{Kim:2012gt}.
The latter mode is quite sensitive to possible beyond-the-SM
contributions; a branching fraction of $\order(10^{-7})$ would
constitute evidence for new physics~\cite{Bar-Shalom:2002sv}. No
experimental information is available for other potential resonance
modes such as $K^{*}(892)^{\pm}\Kmp$, $K^{*}_{0}(1430)^{\pm}\Kmp$,
and $f_0(980)\piz$. For the decay $\Bz\to K^*(892)^{\pm}K^{\mp}$
dominated by internal $W$ exchange [Fig.~\ref{fig:Feyn_diag}(b)],
the branching fraction is predicted to be in the range $10^{-8}$
to $10^{-7}$~\cite{Du:2002up,Beneke:2003zv,Guo:2006uq}.

Another motivation for the study of $\Bz\to\Kp\Km\piz$ comes
from the observation of $\Bu\to\Kp\Km\pip$ by the $\babar$
Collaboration~\cite{Aubert:2007xb}. In particular, an unexpected
structure is seen near $1.5\gevcc$ in the $\Kp\Km$ invariant-mass
spectrum, which accounts for about half of the total events.
Similar structures have also been observed in the Dalitz plots
of $\Bp\to\Kp\Km\Kp$ and $\Bz\to\Kp\Km\Kz$ 
decays~\cite{Garmash:2004wa,Nakahama:2010nj,Lees:2012kxa}. If
these structures are due to a particular $\Kp\Km$ resonant state,
it should show up in $\Bz\to\Kp\Km\piz$; on the other hand, if it
is a reflection from the $b\to d$ penguin, it will not contribute
to $\Kp\Km\piz$. Since the $u$ and $d$  quarks are spectators in
the $b\to u$ tree diagram [Fig.~\ref{fig:Feyn_diag}(a)] for $\Bu\to
\Kp\Km\pip$ and $\Bz\to \Kp\Km\piz$, respectively, one can estimate
the branching fraction for the latter using the $\babar$ results.
Assuming isospin symmetry and the $b\to u$ transition to be the
main contributor to $\Bz\to\Kp\Km\piz$, we expect its branching
fraction to be at the level of $3\times 10^{-6}$, which is well
within Belle's reach. 

Our results are based on a data sample containing $772\times 10^6$
$\BB$ pairs collected at the $\Y4S$ resonance with the Belle
detector~\cite{Belle} at the KEKB asymmetric-energy $e^+e^-$ ($3.5$
on $8.0\gev$) collider~\cite{KEKB}. The principal detector components
used in the study are a silicon vertex detector, a $50$-layer central
drift chamber (CDC), an array of aerogel threshold Cherenkov counters
(ACC), a barrel-like arrangement of time-of-flight scintillation
counters (TOF), and a CsI(Tl) crystal electromagnetic calorimeter
(ECL). All these components are located inside a $1.5$\,T solenoidal
magnetic field.

To reconstruct $\Bz\to\Kp\Km\piz$ decay candidates, we combine two
oppositely charged kaons with a $\pi^0$ meson. Each track candidate
must have a minimum transverse momentum of $100\mevc$, and a distance
of closest approach with respect to the interaction point of less than
$0.2\cm$ in the transverse $r$--$\phi$ plane and less than $5.0\cm$
along the $z$ axis, where the $z$ axis is defined by the direction
opposite the $e^+$ beam. Identification of charged kaons is based on
a likelihood ratio $R_{K/\pi}=\frac{{\cal L}_K}{{\cal L}_K+{\cal L}_\pi}$, 
where ${\cal L}_K$ and ${\cal L}_\pi$ denote the individual likelihoods
for kaons and pions, respectively, calculated using specific ionization
in the CDC, time-of-flight information from the TOF, and the number of
photoelectrons from the ACC. A requirement $R_{K/\pi}>0.6$ is applied
to select both kaon candidates. The kaon identification efficiency is
approximately $86\%$ and the probability of misidentifying a pion as
a kaon is $11\%$. We reconstruct $\pi^0$ candidates from photon pairs
that have an invariant mass between $112$ and $156\mevcc$, corresponding
to $\pm 3.5\sigma$ around the nominal $\pi^0$ mass~\cite{PDG}. These
photons are reconstructed from neutral clusters in the ECL with energy
above $60$ ($100$) $\mev$ in the barrel (endcap) region. In addition,
requirements on the $\pi^0$ decay helicity angle, $|\cos\theta_{\rm hel}|
<0.95$, and the $\pi^0$ mass-constrained fit statistic, $\chi^2_{\rm mass}
<50$, are imposed. Here, $\theta_{\rm hel}$ is the angle between one
of the daughter photons and the $B$ momentum in the $\piz$ rest frame.

$B$-meson candidates are identified using two kinematic variables:
beam-energy constrained mass, $\mbc=\sqrt{E^2_{\rm beam}-\left|\sum_{i}
\vec{p}_i\right|^2}$, and energy difference, $\DeltaE=\sum_{i}E_{i}-
E_{\rm beam}$, where $E_{\rm beam}$ is the beam energy, and $\vec{p}_i$
and $E_i$ are the momentum and energy, respectively, of the $i$th
daughter of the reconstructed $B$ in the center-of-mass (CM) frame.
We retain events with $5.271\gevcc<\mbc<5.289\gevcc$ and $-0.30\gev
<\DeltaE<0.15\gev$ for further analysis. The $\mbc$ requirement
corresponds to approximately $\pm 3\sigma$ around the nominal $\Bz$
mass~\cite{PDG}; we apply a looser window of ($-12\sigma$,\,$+6\sigma$)
around $\DeltaE=0$ because it is used in the fitter (as described
below). The average number of $B$ candidates found per event is $1.3$.
In events with multiple $B$ candidates, we choose the one(s) whose
$\piz$ has the lowest $\chi^2_{\rm mass}$ value. If more than one $B$
candidate shares the same $\pi^0$ meson, the candidate yielding the
best $\Bz$ vertex fit is selected.

The dominant background is from the $e^+e^-\to\qqbar$ ($q=u,d,s,c$)
continuum process. To suppress this background, observables based on
the event topology are utilized. The event shape in the CM frame is
more spherical for $\BB$ events and jetlike for continuum events.
We employ a neural network~\cite{neurobayes} to combine the following
six input variables: the Fisher discriminant formed from $16$ modified
Fox-Wolfram moments~\cite{KSFW}, the cosine of the angle between the
$B$ momentum and the $z$ axis, the cosine of the angle between the $B$
thrust and the $z$ axis, the cosine of the angle between the thrust
axis of the $B$ candidate and that of the rest of the event, the ratio
of the second- to the zeroth-order Fox-Wolfram moments (all of these
quantities being calculated in the CM frame), and the separation along
the $z$ axis between the vertex of the $B$ candidate and that of the
remaining tracks. The training and optimization of the neural network
are accomplished with signal and $\qqbar$ Monte Carlo (MC) simulated
events. The signal MC sample is generated with the {\textsc EvtGen}
program~\cite{evtgen} by assuming a three-body phase space. We require
the neural network output ($\nb$) to be above $0.2$ to substantially
reduce the continuum background. The relative signal efficiency due
to this requirement is approximately $88\%$, whereas the continuum
suppression achieved is close to $92\%$. The remainder of the $\nb$
distribution peaks strongly near $1.0$ for signal, and thus we have
difficulty in modeling it with an analytic function. However, its
transformed variable
\begin{eqnarray}
\nbprim=\log\left[\frac{\nb-\nbmin}{\nbmax-\nb}\right], 
\end{eqnarray}
where $\nbmin=0.2$ and $\nbmax=1.0$, has a distribution with a
Gaussian-like tail.

The background due to $B$ decays via the dominant $b\to c$ transition 
is studied with an MC sample of a collection of such decays. The 
resulting $\mbc$ distribution is found to peak strongly in the signal 
region. We also observe two peaks in the $\Kp\Km$ invariant-mass spectrum
that corresponds to the contributions from (a) $D^0\to\Kp\Km$ peaking
at the nominal $D^0$ mass~\cite{PDG}, and (b) $D^0\to\Km\pip$ with
the peak shifted slightly from the $D^0$ mass owing to $K$--$\pi$
misidentification. To suppress these peaking contributions, we exclude
candidates for which the invariant mass of the $\Kp\Km$ system lies
in the range of $[1846,1884]\mevcc$ (about $\pm 5\sigma$ around the
nominal $D^0$ mass). In the case of (b), we use the pion hypothesis
for one of the tracks. The surviving events constitute the ``generic
$\BB$'' background.

There are a few background modes that contribute in the $\mbc$ signal
region having the $\DeltaE$ peak shifted to positive values. The 
so-called ``rare peaking'' background modes arising mostly from
$K$--$\pi$ misidentification are identified with a $\BB$ MC sample
in which one of the $B$ mesons decays via $b\to u,d,s$ transitions
with known or estimated branching fractions. The rare peaking
background includes the $\Bz\to\Kp\pim\piz$ nonresonant decay as
well as possible intermediate resonant modes that result in the
$\Kp\pim\piz$ final state, such as $\Bz\to\Kstar(892)^0\piz$ and
$\Bz\to\Kstar(892)^+\pim$. The events that remain after removing the
signal and rare peaking components comprise the ``rare combinatorial''
background.

The signal yield is obtained with an unbinned extended maximum likelihood
fit to the two-dimensional distributions of $\DeltaE$ and $\nbprim$. We
define a probability density function (PDF) for each event category $j$
(signal, $\qqbar$, generic $\BB$, rare peaking, and rare combinatorial
$\BB$ backgrounds):
\begin{eqnarray}
{\cal P}^i_j\equiv{\cal P}_j(\DeltaE^{\,i}){\cal P}_j(\nb'^{\,i}),
\end{eqnarray}
where $i$ denotes the event index. Since the correlation between $\DeltaE$
and $\nbprim$ is found to be negligible, the product of two individual PDFs
is a good approximation for the combined PDF. We apply a tight requirement
on $\mbc$ rather than including it in the fitter because it exhibits an
irreducible correlation with $\DeltaE$ owing to shower leakage in the ECL.
The extended likelihood function is
\begin{eqnarray}
{\cal L}=\exp\left(-\sum\limits_j n_j\right)\times 
\prod\limits_i\left[\sum\limits_j n_j{\cal P}^i_j\right],
\end{eqnarray}
where $n_j$ is the yield of event category $j$. The correctly
reconstructed (CR) and misreconstructed fragments of the $B$-meson
decay referred to as self-crossfeed (SCF) components of the signal
are considered distinct in the fitter: their combined PDF is
$n_{\rm sig}\times[(1-f)\,{\cal P}_{\rm CR}+f\,{\cal P}_{\rm SCF}]$,
where $n_{\rm sig}$ is the total signal yield and $f$ is the SCF
fraction fixed to the MC expected value of $3\%$.

\begin{table}[htb]
\centering
\caption{List of PDFs used to model the $\DeltaE$ and $\nbprim$
 distributions for various event categories. G, AG, CB, and Poly2
 denote Gaussian, asymmetric Gaussian, Crystal Ball~\cite{CrystalBall},
 and second-order Chebyshev polynomial function, respectively.}
\label{tab:pdf-shape}
\begin{tabular}{lcccccc}
\hline\hline
Event category & & & $\DeltaE$ & & & $\nbprim$ \\
\hline
CR signal & & & CB+AG & & & 3\,AG \\
SCF signal & & & histogram & & & histogram \\
Continuum $\qqbar$ & & & Poly2 & & & AG \\
Generic $\BB$ & & & Poly2 & & & AG \\
Rare peaking $\BB$ & & & 2\,G & & & AG \\
Rare combinatorial $\BB$ & & & histogram & & & 3\,AG \\
\hline\hline
\end{tabular}
\end{table}

Table~\ref{tab:pdf-shape} lists the PDF shapes used to model the
$\DeltaE$ and $\nbprim$ distributions for each event category.
Distributions that are difficult to parametrize analytically are
modeled with histograms. The yields for all event categories except
the rare peaking $\BB$ background are allowed to vary in the fit.
We fix the yield of the rare peaking $\BB$ component to the value
calculated using the branching fraction measured in an amplitude
analysis of $\Bz\to\Kp\pim\piz$~\cite{BABAR:2011ae}. The following 
PDF shape parameters of the $\qqbar$ background are floated: the
two parameters of the second-order Chebyshev polynomial used for
$\DeltaE$, and the mean and two widths of the asymmetric Gaussian
function used to model $\nbprim$. The PDF shapes for signal and
other background components are fixed to the corresponding MC
expectations. We adjust the parameters of the signal $\DeltaE$ 
and $\nbprim$ PDFs to account for possible data-MC differences,
according to the values obtained with a large-statistics control
sample of $\Bu\to\Dzb(\Kp\pim\piz)\pip$. The same correction
factors are also applied for the rare peaking $\BB$ background.

\begin{figure}[!htb]
\begin{center}
\includegraphics[width=.492\columnwidth]{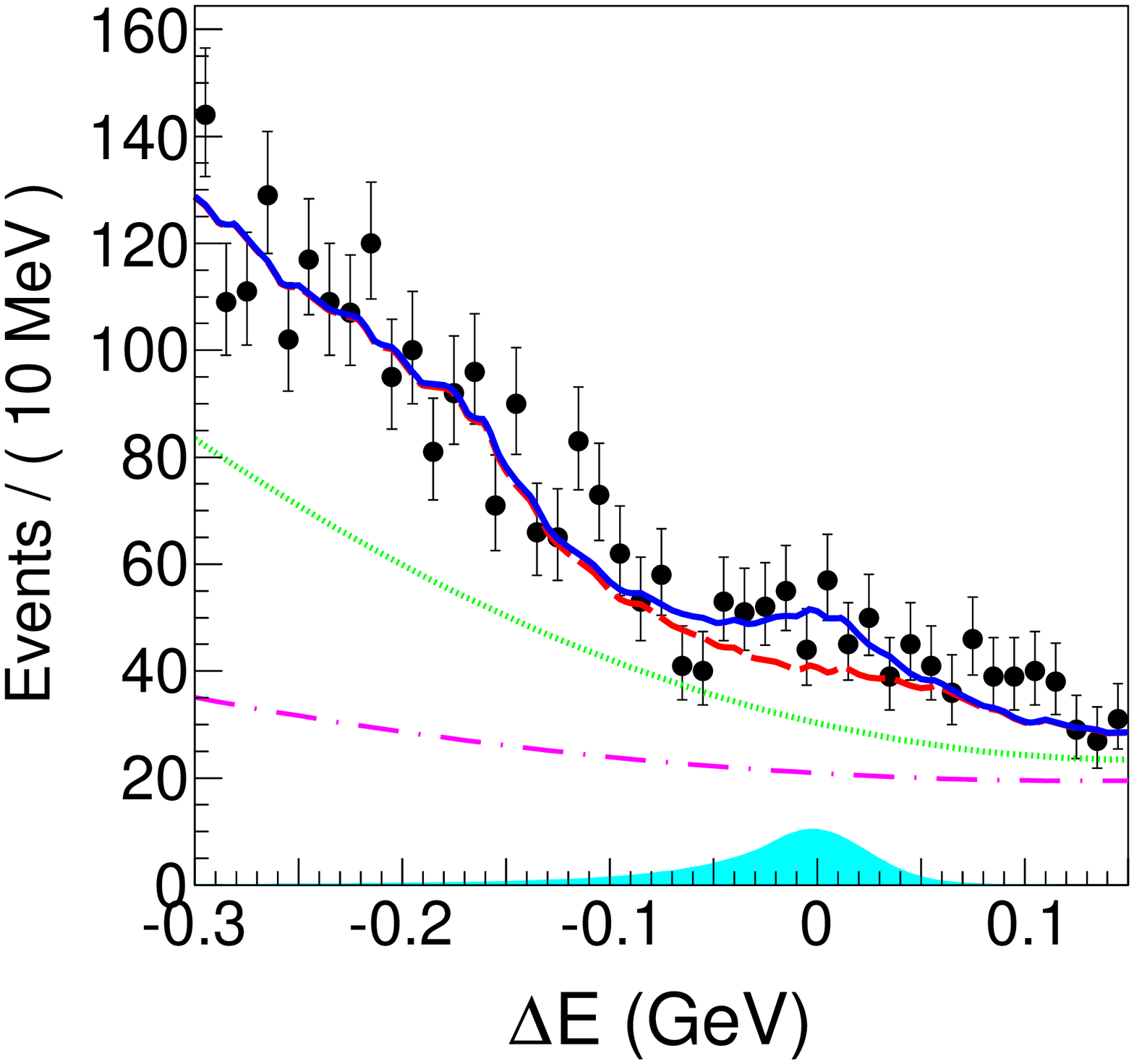}
\includegraphics[width=.492\columnwidth]{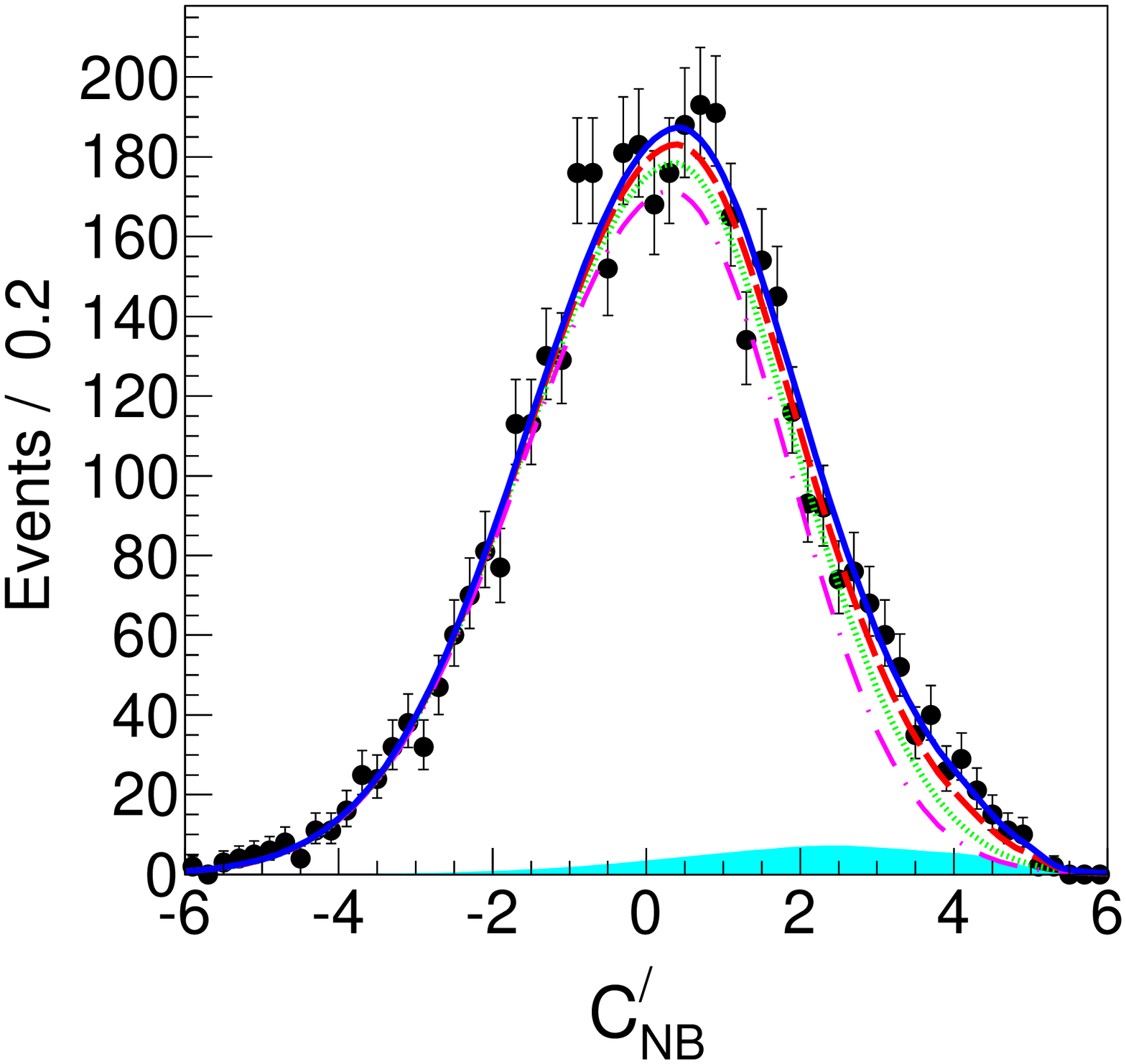}
\end{center}
\caption{(color online). Projections of candidate events onto (left)
 $\DeltaE$ for $\nbprim>3$ and (right) $\nbprim$ for $|\DeltaE|<30\mev$.
 Points with error bars are the data, solid (blue) curves are the
 total PDF, dashed (red) curves are the total background, dotted (green)
 curves are the sum of continuum $\qqbar$ and generic $\BB$ backgrounds,
 dash-dotted (magenta) curves are the continuum $\qqbar$ background,
 and filled (cyan) regions show the signal.}
\label{fig:2D}
\end{figure}

Figure~\ref{fig:2D} shows the $\DeltaE$ and $\nbprim$ projections of
the fit applied to $39066$ candidate events. We obtain $299\pm83$
signal events ($n_{\rm sig}$), $32167\pm428$ continuum $\qqbar$,
$3814\pm517$ generic $\BB$, and $2691\pm321$ rare combinatorial $\BB$
background events. The statistical significance of the signal is
$3.8$ standard deviations. It is calculated as $\sqrt{2\log({\cal L}_0/
{\cal L}_{\rm max})}$, where ${\cal L}_0$ and ${\cal L}_{\rm max}$ are
the fit likelihood values with the signal yield set to zero and the
best-fit case, respectively. The obtained background yields are consistent
with the respective MC predictions. The signal decay branching fraction
is calculated as
\begin{eqnarray}
{\cal B}(\Bz\to\Kp\Km\piz)=
\frac{n_{\rm sig}}{N_{\BB}\times\varepsilon _{\rm rec}\times r_{K/\pi}},
\label{eq:bf}
\end{eqnarray}
where $N_{\BB}$ is the total number of $\BB$ pairs ($772\times 10^6$),
$\varepsilon_{\rm rec}$ is the signal reconstruction efficiency ($19.6\%$)
obtained in the study described below, and $r_{K/\pi}$ denotes the
kaon-identification efficiency correction factor that accounts for 
a small data-MC difference. It is given by
\begin{eqnarray}
r_{K/\pi}\equiv\varepsilon^{\rm data}_{K/\pi}/\varepsilon^{\rm MC}_{K/\pi}, 
\label{eq:rkpi}
\end{eqnarray}
where $\varepsilon^{\rm data}_{K/\pi}$ ($\varepsilon^{\rm MC}_{K/\pi}$) is
the efficiency of the $R_{K/\pi}$ requirement in data (MC simulations). The
$r_{K/\pi}$ value per kaon track is $0.95$, resulting in a total $r_{K/\pi}
=0.95^2=0.90$ for two kaons. We have verified that the $R_{K/\pi}$
correction factor is almost constant over the Dalitz plot. For the
branching fraction calculation presented in Eq.~(\ref{eq:bf}), we assume
equal production of $\Bz\Bzb$ and $\Bp\Bm$ pairs at the $\Y4S$ resonance.
The resulting value is
\begin{eqnarray}
{\cal B}(\Bz\to\Kp\Km\piz)=[2.17\pm0.60\pm0.24]\times 10^{-6},
\label{eq:bfvalue}
\end{eqnarray}
where the uncertainties are statistical and systematic, respectively.
The contributions to the systematic uncertainty are discussed below
and listed in Table~\ref{tab:syst2}.

The uncertainties due to the PDF shape parameters are estimated by varying
all fixed parameters by $\pm1\sigma$. To assign a systematic error for
the histogram PDF used to model $\DeltaE$ for the rare combinatorial
component, we carry out a series of fits by fluctuating each of the
histogram bin contents according to the Poisson distribution. The spread
of the fitted signal yields is taken as the systematic error. We also vary
the yield of final states that dominantly contribute to that component
according to their errors. As we use a fairly complex function (a sum of
three asymmetric Gaussians) to model the signal $\nbprim$ PDF shape, we
evaluate possible systematics due to the uncertainty in the functional
dependence by checking other alternatives. This systematic contribution is
denoted as ``Signal $\nbprim$ functional dependence'' in Table~\ref{tab:syst2}.
The uncertainty due to the fixed (small) SCF fraction is estimated without
knowing {\it a priori} how these SCF events vary across the Dalitz plot.
We adopt a conservative approach to vary the SCF fraction by $\pm50\%$
when calculating the associated systematic error. The potential fit
bias is evaluated by performing an ensemble test comprising $200$
pseudoexperiments, where the signal and rare peaking background
components are embedded from the corresponding MC samples, and the PDF
shapes are used to generate the data for the other event categories. We
obtain an almost Gaussian pull distribution of unit width, and add the
mean and error on the pull in quadrature for assigning the systematics. 
Uncertainty due to continuum suppression is derived with the control
sample by comparing the nominal fit result with that obtained without
any $\nb$ requirement. We estimate the error due to the $\mbc$ requirement
by varying its nominal selection threshold by the resolution. The $\Dstarp
\to\Dz(\Km\pip)\pip$ control sample is used to determine the systematic
uncertainty due to the $R_{K/\pi}$ requirement. The systematic uncertainty
due to $\piz$ reconstruction is evaluated by comparing data-MC differences
of the yield ratio between $\eta\to\piz\piz\piz$ and $\eta\to\pip\pim\piz$.
We use partially reconstructed $\Dstarp\to\Dz(\KS\pip\pim)\pip$ decays to
assign the systematic uncertainty due to charged-track reconstruction
($0.35\%$ per track). To account for the possible variation of efficiency
across the Dalitz-plot distribution, we calculate a weighted signal
reconstruction efficiency by fitting different regions of that distribution.
The mean value is used to obtain the branching fraction and the error is
taken as the systematic contribution due to the efficiency variation. The
total systematic uncertainty is calculated by summing all these uncertainties
in quadrature. To determine the significance of our measurement, we use a
convolution of the statistical likelihood with a Gaussian function of width
equal to the additive systematic errors that only affect the signal yield.
The total significance, including these uncertainties, is $3.5$ standard
deviations.

\begin{table}[htb]
\centering
\caption{Summary of various systematic uncertainties. The first
and second horizontal blocks denote the additive and multiplicative
systematic uncertainties, respectively.}
\label{tab:syst2}
\begin{tabular}{lccc}
\hline\hline
Source & & \multicolumn{2}{c}{Uncertainties ($\%$)} \\
\hline
Signal PDF & & $+3.4$ & $-2.9$ \\
Generic $\BB$ PDF & & $+2.4$ & $-3.1$ \\
Combinatorial background PDF & & $+1.3$ & $-2.0$ \\
Peaking background PDFs & & $+1.7$ & $-1.9$ \\
Fixed histogram PDF & & $+1.7$ & $-2.0$ \\
Signal $\nbprim$ functional dependence & & $+2.3$ & $-2.3$ \\
Fixed SCF fraction & & $+1.7$ & $-1.7$ \\
Fit bias & & $+2.4$ & $-2.4$ \\
Continuum suppression & & $+2.2$ & $-2.2$ \\
Requirement on $\mbc$ & & $+1.5$ & $-0.2$ \\
\hline
Kaon ID requirement & & $+1.9$ & $-1.9$ \\
$\piz$ detection efficiency & & $+4.0$ & $-4.0$ \\
Charged track reconstruction & & $+0.7$ & $-0.7$ \\
Efficiency variation over Dalitz plot & & $+7.5$ & $-7.5$ \\
Number of $\BB$ pairs & & $+1.4$ & $-1.4$ \\
\hline
Total & & $+11.1$ & $-11.3$ \\
\hline\hline
\end{tabular}
\end{table}    

\begin{figure}[!htb]
\begin{center}
\includegraphics[width=.492\columnwidth]{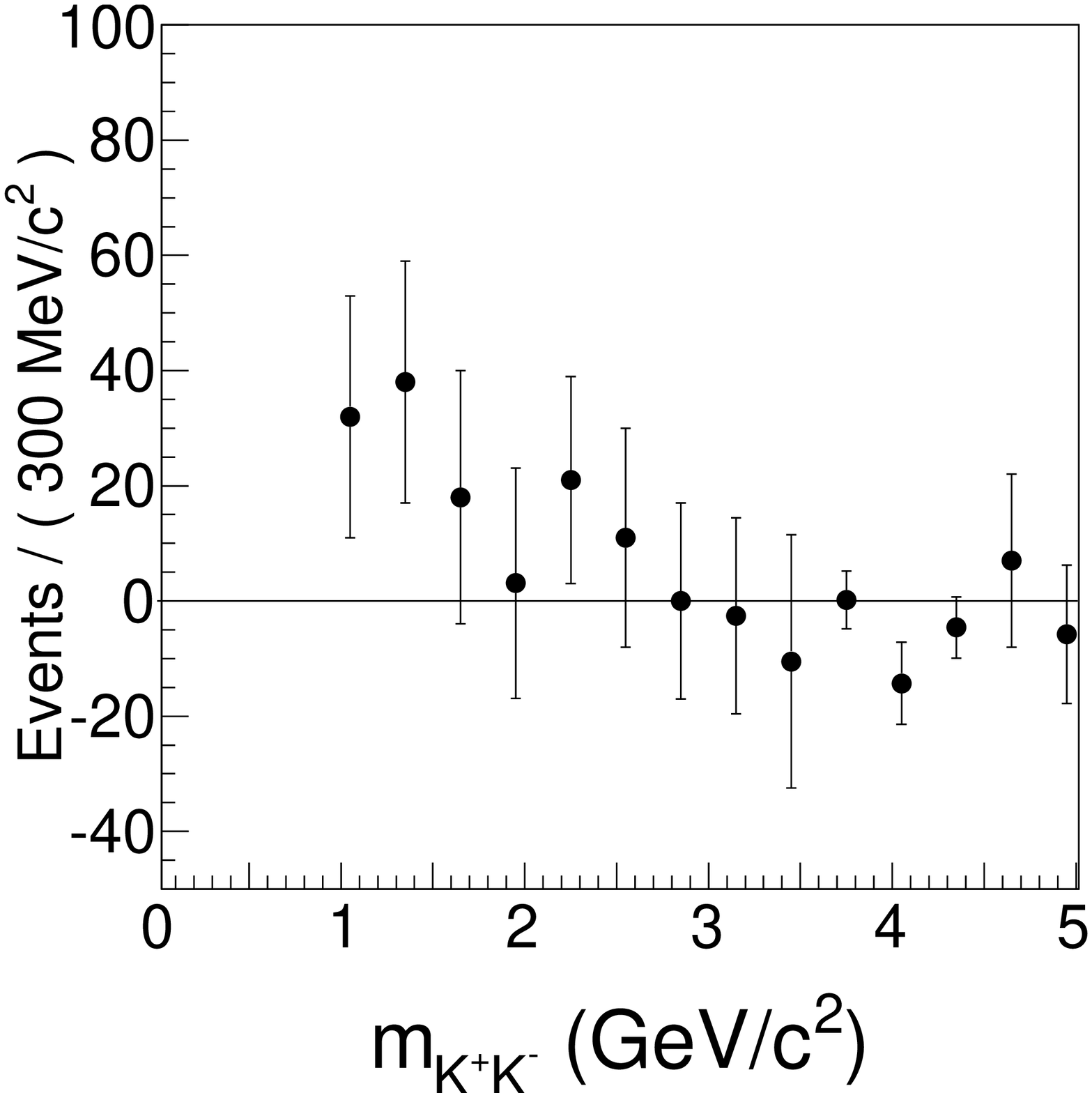}
\includegraphics[width=.492\columnwidth]{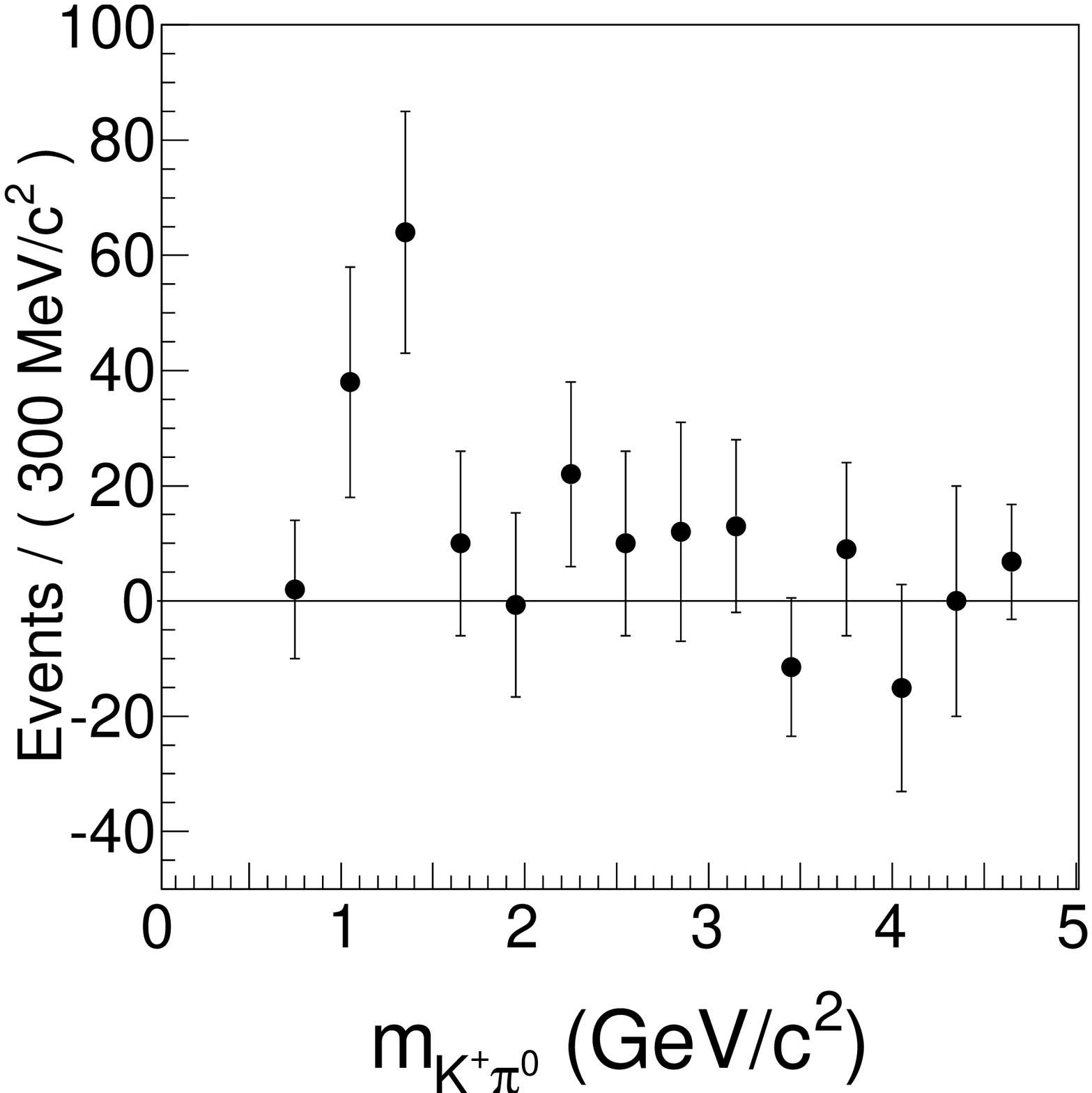}
\end{center}
\caption{Signal yield distributions as a function of (left) $m(\Kp\Km)$ with
$m(\Kp\piz)>1.5\gevcc$ and (right) $m(\Kp\piz)$ with $m(\Kp\Km)>2.0\gevcc$.
Each point is obtained from a two-dimensional [$\DeltaE,\nbprim$] fit.}
\label{fig:mass-fit}
\end{figure}

To elucidate the nature of the observed signal, especially whether there
are contributions from the decays with intermediate resonant states, we
study the $\Kp\Km$ and $\Kp\piz$ invariant mass distributions. We perform
the [$\DeltaE,\nbprim$] two-dimensional fit in bins of the $m(\Kp\Km)$
and $m(\Kp\piz)$ distributions after applying the orthogonal requirements
$m(\Kp\piz)> 1.5\gevcc$ and $m(\Kp\Km)>2.0\gevcc$, respectively. These
requirements suppress kinematic reflections. Figure~\ref{fig:mass-fit}
shows the resulting signal yields along with their statistical errors.
With these data, we cannot make any definitive statement about possible
intermediate $\Kp\Km$ resonances, including the structure seen by $\babar$
near $1.5\gevcc$~\cite{Aubert:2007xb}. It is worth noting here that the 
recent LHCb study of $B^{\pm}\to\Kp\Km\pi^{\pm}$ decays~\cite{lhcb} has
revealed an unidentified structure in the same mass range; however, it
is only present in $\Bp$ events, giving rise to a large local $\CP$
asymmetry. Furthermore, we observe some excess of events around
$1.4\gevcc$ in the $\Kp\piz$ invariant-mass spectrum. A detailed
interpretation will require an amplitude analysis with higher statistics
that would be available at a next-generation flavor factory~\cite{belle2}.

In summary, we report measurement of the suppressed decay $\Bz\to\Kp\Km\piz$
using the full $\Y4S$ data sample collected with the Belle detector. We employ 
a two-dimensional fit for extracting the signal yield. Our measured branching 
fraction ${\cal B}(\Bz\to\Kp\Km\piz)=[2.17\pm0.60\stat\pm0.24\syst]\times
10^{-6}$ constitutes the first evidence for the decay.

We thank the KEKB group for the excellent operation of the
accelerator, the KEK cryogenics group for the efficient
operation of the solenoid, and the KEK computer group,
the National Institute of Informatics, and the 
PNNL/EMSL computing group for valuable computing
and SINET4 network support.  We acknowledge support from
the Ministry of Education, Culture, Sports, Science, and
Technology (MEXT) of Japan, the Japan Society for the 
Promotion of Science (JSPS), and the Tau-Lepton Physics 
Research Center of Nagoya University; 
the Australian Research Council and the Australian 
Department of Industry, Innovation, Science and Research;
Austrian Science Fund under Grant No. P 22742-N16;
the National Natural Science Foundation of China under Contracts
No.~10575109, No.~10775142, No.~10875115, and No.~10825524; 
the Ministry of Education, Youth and Sports of the Czech 
Republic under Contract No.~MSM0021620859;
the Carl Zeiss Foundation, the Deutsche Forschungsgemeinschaft
and the VolkswagenStiftung;
the Department of Science and Technology of India; 
the Istituto Nazionale di Fisica Nucleare of Italy; 
The BK21 and WCU program of the Ministry Education Science and
Technology, National Research Foundation of Korea Grants
No.~2010-0021174, No.~2011-0029457, No.~2012-0008143, No.~2012R1A1A2008330,
BRL program under NRF Grant No. KRF-2011-0020333,
and GSDC of the Korea Institute of Science and Technology Information;
the Polish Ministry of Science and Higher Education and 
the National Science Center;
the Ministry of Education and Science of the Russian
Federation and the Russian Federal Agency for Atomic Energy;
the Slovenian Research Agency;
the Basque Foundation for Science (IKERBASQUE) and the UPV/EHU under 
program UFI 11/55;
the Swiss National Science Foundation; the National Science Council
and the Ministry of Education of Taiwan; and the U.S.\
Department of Energy and the National Science Foundation.
This work is supported by a Grant-in-Aid from MEXT for 
Science Research in a Priority Area (``New Development of 
Flavor Physics''), and from JSPS for Creative Scientific 
Research (``Evolution of Tau-lepton Physics'').


\begin{thebibliography}{99}
\bibitem{charge}
  Throughout this paper, inclusion of the charge-conjugate
  process is implied unless otherwise explicitly mentioned.
\end{thebibliography}
\end{document}